\documentstyle[aps,preprint,tighten,epsf,floats]{revtex}
\newcommand\beq{\begin{equation}}
\newcommand\eeq{\end{equation}}
\newcommand\bqa{\begin{eqnarray}}
\newcommand\eqa{\end{eqnarray}}

\def\square{\vcenter{\vbox{\hrule height.4pt
          \hbox{\vrule width.4pt height8pt
          \kern8pt\vrule width.4pt}\hrule height.4pt}}}

\def\gsim{\stackrel{>}{\sim}}

\begin{document}
\draft

\title{The Dilute Bose-Einstein Condensate
with Large Scattering Length}

\author{Eric Braaten,
        H.-W. Hammer, and
        Thomas Mehen}
\address{Department of Physics, The Ohio State University,
Columbus, OH 43210, USA}

\date{\today}

\maketitle

\begin{abstract}
We study a dilute Bose gas of atoms whose scattering length $a$ is large
compared to the range of their interaction. We calculate the energy
density $\cal{E}$ of a homogeneous Bose-Einstein condensate (BEC) to
second order in
the low-density expansion, expressing it in terms of $a$ and a second
parameter $\Lambda_*$ that determines the low-energy observables in the
3-body sector. The second-order correction to $\cal{E}$
has a small imaginary part that reflects the instability due to
3-body recombination. In the case of a trapped BEC
with large negative $a$, we calculate the coefficient of
the 3-body mean-field term in $\cal{E}$ in terms of $a$ and
$\Lambda_*$. It can be very large if there is an Efimov state near
threshold.                                                         
\end{abstract}

\bigskip

\thispagestyle{empty}
\newpage
\setcounter{page}{1}

Bose-Einstein condensates of atoms have been extensively studied both
experimentally and theoretically for 2 cases:  liquid helium
\cite{Helium} and dilute vapors of alkali atoms \cite{Leggett-01}.  The
low-energy length scale $\ell = (mC_6/ \hbar^2)^{1/4}$ set by the van
der Waals interaction $-C_6/r^6$ is the natural length scale for the
S-wave scattering length $a$ and other parameters
in the low-energy expansions of scattering amplitudes.
In the case of liquid helium,
the interparticle spacing $n^{-1/3} = 3.6$ \AA \ is much smaller than
$\ell = 37$ \AA \ and $a=104$ \AA \cite{Gri-00}.
Because the diluteness variable $\sqrt{n \ell^3} \approx 33$
is much greater than 1, the properties of the
Bose-Einstein condensate (BEC) in liquid helium
depend on the detailed behavior of the interatomic potential.
In contrast, BEC's consisting of dilute gases of
atoms with $n \ell^3 \ll 1$ have  universal properties that depend on
the interatomic potential only through the single low-energy parameter
$a$. For example, the first few terms in
the low-density expansions for the energy density and
the condensate fraction of a homogeneous condensate
can be calculated as expansions in powers of $\sqrt{na^3}$.

A fundamental open problem is the behavior of a dilute
BEC ($n \ell^3 \ll 1$)
with large scattering length $(a \gg
\ell)$ when $na^3$ is comparable to or much greater than 1. The
diluteness condition $n \ell^3 \ll 1$ excludes the case of liquid
helium. The condition $na^3 \gsim 1$ implies that the low-density
expansion is not applicable, because the expansion parameter $\sqrt{16
\pi na^3}$ is not small. The most basic question is whether a
BEC even exists as a well-defined quasistable state
in this limit. If so, does it have any universal properties that depend
on the interatomic potential only through the scattering length $a$?
These questions can be investigated experimentally in vapors of alkali
atoms by tuning a background magnetic field to a Feshbach resonance
\cite{Feshbach}. They can also be studied using numerical methods
\cite{GBC-99,CHMMPP}.

In this letter, we take a small step towards addressing this problem by
studying the homogeneous BEC with large scattering
length $(a \gg \ell)$ in the extremely dilute limit $na^3 \ll 1$. We
calculate its energy density to second order in $\sqrt {na^3}$,
determining the coefficient of $na^3$ in terms of a low-energy parameter
$\Lambda_*$. The parameter $\Lambda_*$ can be determined from any 
low-energy 3-body observable, e.g. the binding energy of the shallowest
3-body bound state. We also determine the coefficient of
the 3-body term in the mean-field contribution to the energy density for
$a < 0$. Finally we comment on the implications of our results for the
case of large $na^3$.

The dilute Bose gas with large scattering length is distinguished from
the generic case by having a 2-body scattering amplitude $f = -a/
(1+iak)$ that has the scale-invariant form $f \approx i/k$ for
wavenumbers in the range $1/a \ll k \ll 1/ \ell$ \cite{M-S-W}. If $a >
0$, it is also characterized by the existence of a 2-body bound state
with binding energy $B_2 = \hbar^2/ma^2$. As first pointed out by Efimov
\cite{Efimov}, the 3-body sector also exhibits universal properties in
the limit of large scattering length, such as the existence
of a large number of 3-body bound states. The number of these Efimov
states is roughly $\ln (|a|/\ell) / \pi$. The Efimov spectrum and
other low-energy 3-body observables involving wavenumbers $k \ll 1/
\ell$ are universal in the sense that they depend on the interatomic
potential only through $a$ and a single 3-body parameter.
A convenient choice for this 3-body parameter is the low-energy
parameter $\Lambda_*$ introduced in Ref.~\cite{BHK-99}.

In the dilute limit $na^3 \ll 1$, the properties of the Bose gas can be
calculated using the low-density expansion. The first few terms in the
low-density expansion of the energy density of a homogeneous
BEC are
\bqa
\cal{E}&=& 2\pi \hbar^2 a n^2 / m \Big\{1 + 128 /(15 \sqrt{\pi})
\sqrt{na^3}
\nonumber\\
&+& 8 (4 \pi -3 \sqrt {3}) / 3 \left[\ln (n a^3)+ 4.72 + 2 B \right]
na^3 + . . . \Big\}.
\label{Edens}
\eqa
The $\sqrt {na^3}$ correction was first calculated by Lee and Yang in
1957 \cite{LY-57}. The coefficient of the logarithm in the $na^3$
correction was calculated in 1959 \cite{Wu-59}. The constant under the
logarithm was calculated by Braaten and Nieto in 1999 \cite{BN-99}.  It
was expressed in terms of an effective 3-body coupling constant  $g_3
(\kappa)$ that can be determined by measuring the low-energy behavior of
the 3-atom elastic scattering rate. This ``running coupling constant"
depends on an arbitrary wavenumber $\kappa$:
\beq
g_3 (\kappa)= 384 \pi (4\pi -3 \sqrt{3})
[\ln (\kappa a) + B] \hbar^2 a^4/m,
\label{g3}
\eeq
where $B$ is the same constant as in (\ref{Edens}). The $na^3$
correction in (\ref{Edens}) is the sum of a mean-field contribution
$g_3 (\kappa) n^3/36$ coming from the effective 3-body contact
interaction and a 2-loop contribution from quantum fluctuations around
the mean field, which also depends on $\kappa$. The dependence on
$\kappa$ cancels in (\ref{Edens}), reflecting the arbitrariness in the
separation of the energy density into mean-field and quantum-fluctuation
contributions. The next nonuniversal term in the low-density
expansion is determined by the effective range for 2-body scattering
\cite{BHH-01}. It is suppressed by $(na^3)^{3/2} \ell /a$, and is
therefore completely negligible in the limit of large scattering length.

To determine the constant $B$ in the expression (\ref{Edens}) for the
energy density for an extremely dilute Bose gas with large scattering
length, we need to calculate $g_3 (\kappa)$ as a function of $a$ and
the low-energy 3-body parameter $\Lambda_*$
introduced in Ref.~\cite{BHK-99}.
This can be accomplished by calculating the T-matrix
element for 3-atom elastic scattering in the low-energy limit. As the
total energy $E$ of the 3 atoms goes to $0$, the T-matrix element is the
sum of divergent terms proportional to $1/E $, $1/ \sqrt {E}$, and $ \ln
E$ \cite{AR-70} and a remainder. The dependence of the T-matrix element
on $\Lambda_*$ enters only through the remainder, which includes a term
$- g_3 (\kappa)$. For the case $a < 0$,  Efimov used simple probability
arguments to deduce the dependence of the  remainder on $\Lambda_*$
\cite{Efimov}. These arguments imply
\beq
B =  b_1 + b_2 \tan (s_0 \ln (|a| \Lambda_*) + \beta) \qquad (a<0),
\label{g3-neg}
\eeq
where $s_0=1.0064$.
Efimov did not determine the functional form of $B$ for $a>0$. However
unitarity requires that it have an imaginary part that is related to the
3-body recombination rate into the shallow 2-body bound state
\cite{Macek-Greene,BBH-00}:
\beq
{\rm Im} \, B   = 0.022 \cos^2 \left(s_0 \ln (a \Lambda_*) + 1.76
\right)\qquad (a>0).
\label{g3-pos}
\eeq
The 3-body recombination rate has zeroes at values of $a \Lambda_*$ that
differ by multiplicative factors of 22.7, and consequently
${\rm Im} \, B$ also vanishes at those points.

A convenient way to calculate $g_3 (\kappa)$, both for $a>0$ and $a<0$,
is to use the effective field theory method developed by Bedaque,
Hammer, and van Kolck in Ref. \cite{BHK-99}. The effective theory is
defined by the lagrangian density
\bqa
{\cal L} &=& i \psi^* {\partial \ \over \partial t} \psi
+{1 \over 2m} \psi^* \nabla^2 \psi
-{2 \pi a \over m} |\psi|^4
\nonumber\\
&& + {2 \pi a \over m} |d- \psi^2|^2 - {g_3 \over 36} |d|^2 |\psi|^2,
\label{Leff}
\eqa
where we have set $\hbar=1$. The auxiliary field $d$, which annihilates
a pair of atoms at a point, can be eliminated using the equation of
motion $d=\psi^2(1 + m g_3/(72 \pi a) |\psi|^2 + ...)$.
However, it is more convenient to keep it, because
3-body observables can be conveniently calculated in terms of 4-point
Green's functions of $\psi, d, \psi^\dagger$ and $d^\dagger$. For
example, the T-matrix element for 3-atom elastic scattering can be
expressed as a sum of 9 connected 4-point Green's functions. If the
incoming atoms have momenta ${\bf k}_1$, ${\bf k}_2$, and ${\bf k}_3$
with ${\bf k}_1 + {\bf k}_2 +{\bf k}_3=0$, the external energies and
momenta of $\psi$ and $d$ are set to $(k_1^2/2m, {\bf k}_1)$ and
$((k_2^2 + k_3^2)/2m, {\bf k}_2 + {\bf k}_3)$, respectively, and
similarly for the outgoing atoms. The Green's function is then summed
over the 3 cyclic permutations of ${\bf k}_1$, ${\bf k}_2$, and ${\bf
k}_3$, and over the 3 cyclic permutations of the final momenta to get
the T-matrix element. The Green's functions can be calculated
nonperturbatively by solving the integral equations shown in Fig.~1a.
The single lines are propagators for $\psi$ and the double lines
are exact propagators for $d$.
An ultraviolet cutoff $\Lambda$ must be imposed on
the loop momentum in the integral equation. 
The 3-body coupling constant $g_3$ in (\ref{Leff}) is tuned
as a function of $\Lambda$ so that the low-energy 3-body observables 
depend only on $a$ and the low-energy parameter $\Lambda_*$.

%%%%%%%%%%%%%%%%%%%%%%%%%%%%%%%%%%%%%%%%%%%%%%%%%%%%%%%%%%%%%%%%%%%%%%%%

\begin{figure}[ht]
\bigskip
\epsfxsize=15.cm
\centerline{\epsffile{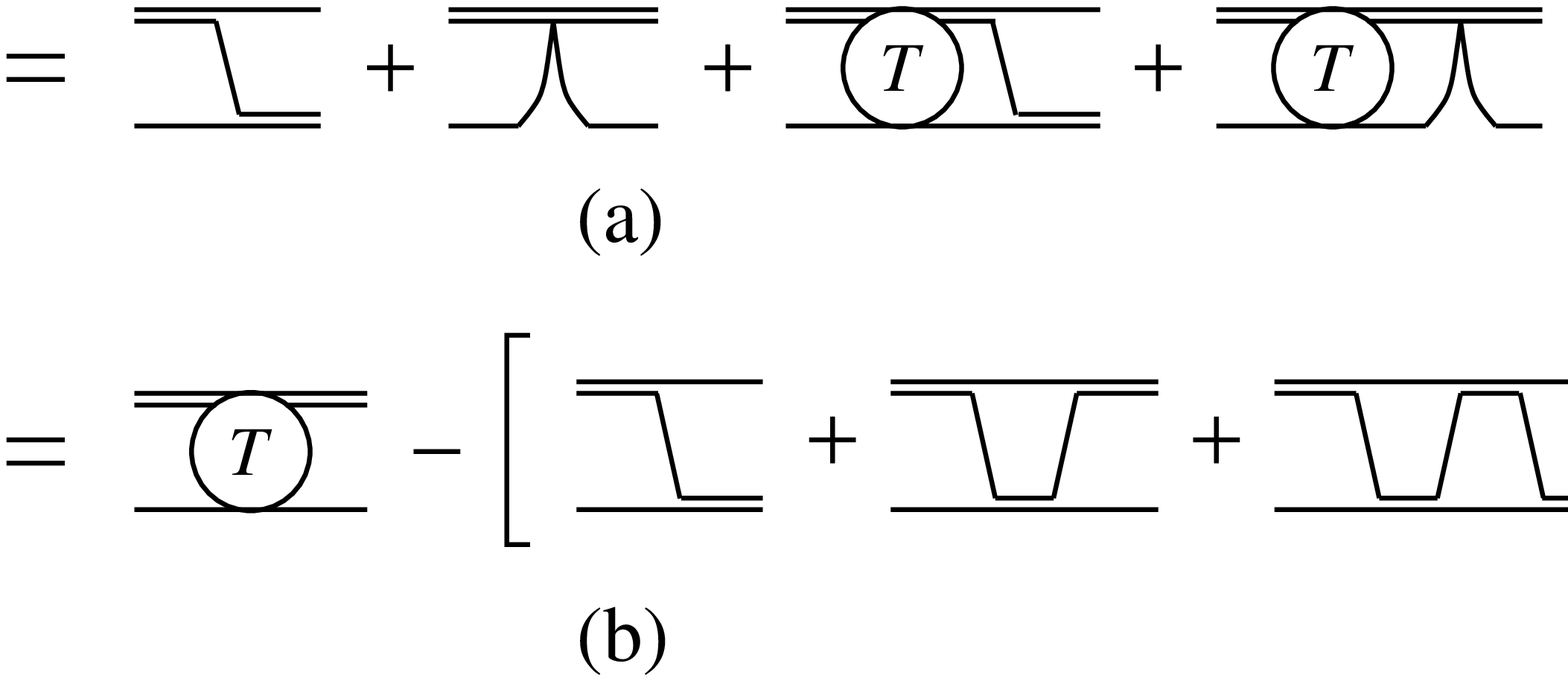}}
\bigskip
\caption{(a) Integral equation satisfied by the off-shell amplitude
$T(p)$. (b) Definition of the subtracted amplitude $\overline{T}(p)$.}
\label{fig2}
\end{figure}

%%%%%%%%%%%%%%%%%%%%%%%%%%%%%%%%%%%%%%%%%%%%%%%%%%%%%%%%%%%%%%%%%%%%%%%%

It is not necessary to calculate a physical observable to determine
$g_3(\kappa)$. For small external energies and momenta $(\omega, {\bf
k})$ with $m|\omega| a^2 \ll 1$ and $|{\bf k}| \ll 1$, a Green's
function associated with the lagrangian (\ref{Leff}) can also be
calculated perturbatively in $a$ and $g_3$ using the methods of
Ref.~\cite{BN-99}. By computing the same Green's function using the
nonperturbative method of Ref.~\cite{BHK-99}, we can determine $g_3$ as
a function of $a$ and $\Lambda_*$. We choose the two
particle-irreducible connected Green's function $T(p)$ with external
energies and momenta $(0, {\bf 0})$ for $\psi$ and $d$, $(p^2/2m, {\bf
p})$ for $\psi^\dagger$, and $(-p^2/2m, -{\bf p})$ for $d^\dagger$. As
$p \to 0$, $T (p)$ has divergent terms proportional to $1/p^2$, $1/p$,
and $\ln(p)$ that come from the diagrams in brackets in Fig.~1b.
The coupling constant $g_3(\kappa)$ in Ref.~\cite{BN-99} was defined
using dimensional regularization in $3-2 \epsilon$ dimensions to
regularize the ultraviolet divergences and minimal subtraction with
renormalization scale $\kappa$ to remove the poles in $\epsilon$.
Calculating the diagrams that contribute to $T (p)$ in the limit $p
\rightarrow 0$ using perturbation theory to first order $g_3$ and 4th
order in $a$, we obtain
\bqa
T(p)&\longrightarrow& - g_3 (\kappa) /36  + 16 \pi^2 a^4 / m\Big\{ 1 /
(p^2 a^2) - 2 \pi / (3 pa)
\nonumber\\
&& - 2  (4 \pi -3 \sqrt{3}) / (3 \pi) \left[\ln (p/ \kappa) + A_{\rm MS}
\right] \Big\} ,
\label{T-dimreg}
\eqa
where $A_{\rm MS} = -0.77268$ and the subscript MS stands for ``minimal
subtraction." The nonperturbative calculation of the Green's function
$T(p)$ using the method of Ref.~\cite{BHK-99} requires solving the
integral equation in Fig.~1a.  It is convenient to define a
subtracted Green function $\overline{T} (p)$ that has a finite limit as
$p \to 0$ using the subtraction shown in Fig.~1b. We solve the integral
equation for $\overline {T} (p)$ as a function of $p, a,$ and
$\Lambda_*$ and then set $p=0$. The resulting expression for $T(p)$ in
the limit $p \rightarrow 0$ is
\bqa
T(p)&\longrightarrow&  16 \pi^2 a^4 / m \Big \{ 1 / (p^2 a^2) - 2 \pi/
(3 pa)
\nonumber\\
&& - 2(4 \pi -3 \sqrt{3}) / (3 \pi) \left[\ln (pa) + A (a \Lambda_*)
\right] \Big \},
\label{T-inteq}
\eqa
where $A (a \Lambda_*)$ is a function of $a \Lambda_*$
that is determined numerically.
Matching the expressions (\ref{T-dimreg}) and
(\ref{T-inteq}), we find that the constant $B$ in (2) is given by $B =
-A_{\rm MS}+A (a \Lambda_*)$.

We first consider the case of large positive scattering length. For
$a>0$, our result for ${B}$ is complex. Its imaginary part is equal to
(\ref{g3-pos}), as required by unitarity.  For the real part,
we obtain a good fit with the empirical formula
\beq
{\rm Re} \, B  =  1.22 + 0.021 \cos^2(s_0 \ln(a \Lambda_*) + 1.0)
\qquad (a>0).
\eeq
The real part is much larger than the imaginary part given in
(\ref{g3-pos}).  The oscillatory term in the real part has almost the same
amplitude as the imaginary part, but a different phase.

In the coefficient of $na^3$ in the expression (\ref{Edens})
for the energy density, the real part of the factor in brackets
can be separated into 2 terms:
$2 \ln (\kappa a) + 2 {\rm Re} B$,  which comes from the
mean-field contribution to the energy density, and ln$(na/ {\kappa^2})+
4.72$, which comes from quantum field fluctuations.  Although $\kappa$
is arbitrary, we can by a suitable choice of $\kappa$ arrange for most
of the $na^3$ correction to come from the mean-field contribution. The
smallest value of $\kappa$ that is physically reasonable is of order
$\sqrt{16 \pi n a}$, which is the inverse of the coherence length. Below
this scale, the dispersion relation for the atoms is modified by
collective effects and 3-atom  scattering is therefore no longer
described accurately by the vacuum T-matrix. If we choose $\kappa = 1.5
\sqrt{16 \pi na}$, the entire $na^3$ correction in (\ref{Edens})
is taken into account by the mean-field contribution $g_3(\kappa) n^3/36$.

The $na^3$ correction to the energy density has an imaginary part that
comes from the imaginary part of $B$ given in (\ref{g3-pos}). This
imaginary part reflects the fact that the BEC is
only a quasistable state.  Atoms will be continually lost from the
condensate by 3-body recombination into the shallow 2-body bound state
and a recoiling atom. In the mean field approximation,
the rate of decrease of the number density is
$(\partial / \partial t) n  = - ({\rm Im} g_3/6 \hbar)n^3$. 
Note that ${\rm Im} g_3$ is independent of $\kappa$. The
energy loss rate is given by $\partial{\cal E}/\partial t=
(\partial{\cal E}/\partial n)\partial n/\partial t$. The
condensate is particularly stable at values of $a$ that correspond to
$\ln (a \Lambda_*) = 2.93$ mod $\pi/s_0$.
At these discrete values of
$a$ that differ by multiplicative factors of 22.7, there is an
interference effect that causes the 3-body recombination rate to vanish
\cite{Macek-Greene,BBH-00}.
The recombination process that gives the energy density an imaginary
part produces atoms and bound states with wavenumber $k = 2/(\sqrt{3}
a)$. These high energy atoms and bound states can subsequently
thermalize and cool by elastic scattering with atoms in the condensate.
The cross section for elastic scattering of the atoms and bound states
can be calculated as a function of $a$ and $\Lambda_*$. For example,
the scattering length was calculated in Ref.~\cite{BHK-99}
and the effective range in Ref.~\cite{HM-01}.

We now turn to the case of large negative scattering length. If $a< 0$,
the expression (\ref{Edens}) for the energy density does not apply,
because the homogeneous condensate is unstable to collapse. However a
condensate with $a<0$ can be stabilized by a trapping potential as long
as the number of atoms is below some critical value 
$N_{\rm max}$ \cite{Rup-95}. 
An effective 3-body contact interaction $-g_3$ gives a mean-field
contribution $g_3  n^3/36$ to the energy density. The effect of such a
term on the stability of a trapped condensate has been studied by adding
a 3-body term to the Gross-Pitaevski equation \cite{GFTC-00}. Even a
small positive value of $g_3$ can considerably increase the critical
number $N_{\rm max}$.

In the case of large negative scattering length, the 3-body coupling
constant $g_3 (\kappa)$ is given by (\ref{g3}). The dependence of our
result for $B$ on $a \Lambda_*$ has the functional form (\ref{g3-neg})
predicted by Efimov.  The values of the constants are
$b_1=1.23$, $b_2 = -3.16$,
and $\beta=0.19$. The appropriate choice for $\kappa$ in this case is
the lowest wavenumber above which 3-body  scattering can be described by
the T-matrix for scattering in the vacuum. The vacuum T-matrix becomes
inaccurate not only because of the nontrivial quasiparticle dispersion
relation but also because of the inhomogeneities of the trapping
potential $V(r)$.  The minimum value of $\kappa$ set by the
coherence length is $\kappa^2 > 4 m (\mu - V(r))$, where $\mu$ is the chemical
potential. For a harmonic potential with frequency $\omega$,
the minimum set by the inhomogeneity of the
potential is $\kappa^2 > m \omega /\hbar$. In
previous studies of the mean-field effects of the 3-body term
\cite{GFTC-00}, the coefficient $g_3$ was assumed to be a constant. If
the large scattering length is obtained by tuning the magnetic field to
a Feshbach resonance, $g_3$ is
given by the expression (\ref{g3}), which scales roughly like $a^4$.
Note that the constant $B$ given in (\ref{g3-neg}) diverges at values of
$a$ that correspond to $\ln (a \Lambda_*) = 1.37 $ mod $\pi/s_0$. At
these discrete values of $a$, there is an Efimov state at the 3-atom
threshold. Near these values of $a$, the 3-body term in the
Gross-Pitaevski equation becomes particularly important. However if the
Efimov state is too close to threshold, the mean-field approximation
breaks down because the energy dependence of the 3-body elastic
scattering amplitude from terms proportional to $1/(E + B_3)$ is not
properly taken into account.

We turn finally to the problem of a homogeneous BEC
with a large scattering length $(a \gg \ell)$ that is dilute enough so
that $n \ell^3 \ll 1$ but dense enough so that $na^3 \gg 1$. Our results
for the extremely dilute limit provide a reference point for discussing
this problem. The fact that our energy density has an imaginary part
serves to emphasize that such a state will at best be quasistable. Monte
Carlo methods that search for the absolute ground state \cite{GBC-99} will
therefore be of limited utility. In the extremely dilute limit, the
condensate is quasistable because the imaginary part of the energy
density is suppressed by $na^3$. Is there any mechanism that can provide
quasistability at large $na^3$? If a quasistable Bose-Einstein
condensate does exist, there are some other obvious questions. At what
rate does the number density in the condensate decay, and what is the
fate of the atoms that disappear from the condensate? We have answered
these questions for the extremely dilute limit.

Assuming that a quasistable BEC exists in the limit
$na^3 \to \infty$, does it have universal properties that depend only on
$a$?  If so, dimensional analysis implies that the energy density must
have the form ${\cal E} = C \hbar^2 n^{5/3}/m$, where $C$ is a constant.
In the extremely dilute limit, we found that ${\cal E}$
depends on $\Lambda_*$ at second order in $\sqrt{na^3}$.
Does ${\cal E}$ depend on $\Lambda_*$ in the limit $na^3 \to \infty$?
If so, the coefficient $C$ in the energy density is not a constant,
but depends on $n$ in a peculiar way. It must be a
periodic function of $\ln (n^{-1/3} \Lambda_*)$ that returns to the same
value when $n$ is increased by a factor of about 11,700. This follows
from the discrete scaling symmetry of low-energy 3-body observables
discovered by Efimov \cite{Efimov,HM-01}, which implies that $C (n^{-1/3}
\Lambda_*)= C (22.7 n^{-1/3} \Lambda_*)$. This discrete scaling symmetry
is related to the fact that there are Efimov states with sizes differing
by factors of 22.7 and ranging all the way from order $\ell$ to order
$a$. The question of whether $C$ depends on $\Lambda_*$ is not addressed
by the constrained variational calculations of Ref.~\cite{CHMMPP}. Their
variational ansatz excludes states with 3-body configurations that are
sensitive to $\Lambda_*$. For example, in the 3-body sector, it would
exclude the Efimov states.

If the coefficient $C$ in the energy density depends on the 3-body
parameter $\Lambda_*$, one might worry that it may also depend on
infinitely many other low-energy parameters associated with 4-body and
higher $n$-body observables. If this is the case, the dilute
BEC with large scattering length would have no
universal properties at large $na^3$. The most favorable possibility
is that in the limit of large scattering length,
the  low-energy observables in the
4-body and higher $n$-body sectors are all calculable to leading order
in $\ell /a$ in terms of $a$ and $\Lambda_*$ only. If this is the case,
then a dilute BEC with
large $na^3$ may have properties that are universal in the sense that
they are determined by $a$ and $\Lambda_*$ only.

This research was supported in part by DOE grant DE-FG02-91-ER4069
and by NSF grant PHY-9800964.

{\it Note added.---}After this work was submitted for publication,
we became aware of a related study by Bulgac \cite{Bul-99}.

\end{document}